\documentclass[11pt]{article}

\usepackage[margin=0.75in]{geometry}
\usepackage{mathpazo}
\usepackage[T1]{fontenc}

\usepackage{amsmath, amssymb, amsthm}
\usepackage{mathtools}

\usepackage{microtype}

\usepackage{authblk}
\usepackage{fancyhdr}
\usepackage{titlesec}
\usepackage{enumitem}
\usepackage{booktabs}
\usepackage{array}

\usepackage{tikz}
\usetikzlibrary{positioning, arrows.meta}

\usepackage{cite}
\usepackage[colorlinks=true,linkcolor=black,citecolor=black,urlcolor=black]{hyperref}

\titleformat{\section}{\normalfont\Large\bfseries}{\thesection}{1em}{}
\titleformat{\subsection}{\normalfont\large\bfseries}{\thesubsection}{1em}{}

\theoremstyle{plain}
\newtheorem{theorem}{Theorem}

\newtheorem{proposition}[theorem]{Proposition}

\theoremstyle{definition}
\newtheorem{definition}[theorem]{Definition}
\newtheorem{example}[theorem]{Example}
\newtheorem{problem}[theorem]{Problem}

\theoremstyle{remark}
\newtheorem{remark}[theorem]{Remark}

\newcommand{\NP}{\mathsf{NP}}
\newcommand{\coNP}{\mathsf{coNP}}

\newcommand{\PSPACE}{\mathsf{PSPACE}}

\newcommand{\oP}{\oplus\mathsf{P}}
\newcommand{\SigTwo}{\Sigma_2^{\oplus\mathsf{P}}}
\newcommand{\PiOne}{\Pi_1^{\oplus\mathsf{P}}}

\newcommand{\BCD}{\textsc{BCD}}
\newcommand{\BCDK}{\textsc{BCD}_{K}}
\newcommand{\Taut}{\textsc{Tautology}}

\newcommand{\F}{\mathbb{F}}

\newcommand{\abs}[1]{\lvert #1 \rvert}

\title{\vspace{-2em}\bfseries
  Boolean Derivative Certificates and Maximal ANF Terms
}
\author{Nicholas Smirnov}
\affil{Department of Computer Science, Stony Brook University}
\date{}

\begin{document}
\maketitle
\thispagestyle{fancy}

\begin{abstract}
\noindent For a Boolean function $f:\{0,1\}^n\to\{0,1\}$, the higher-order Boolean derivative $D_S f$ computes the parity of $f$ over each $S$-dimensional subcube. We prove that $D_S f\equiv 1$ exactly when $S$ is a maximal monomial support in the algebraic normal form of $f$. This correspondence motivates the derivative certificate depth $\Delta_\partial(f)$, defined as the minimum degree of a nonempty maximal algebraic normal form monomial. We study the decision problem $\BCD$: given $f$ and $k$, decide whether $\Delta_\partial(f)\le k$. We show that this problem is $\coNP$-complete for every fixed constant $k\ge 1$, $\NP$-complete when $k=n$, and belongs to $\Sigma_2^{\oplus P}$ when $k$ is part of the input; in the variable-$k$ case it is also both $\NP$-hard and $\coNP$-hard. These results identify the $\exists\forall\oplus$ upper bound and leave $\Sigma_2^{\oplus P}$-completeness as an open problem.
\end{abstract}

\section{Introduction}

The \emph{algebraic normal form} (ANF) of a Boolean function $f$, its unique representation as an $\F_2$-multilinear polynomial, has been quite foundational to Boolean function theory since Zhegalkin~\cite{zhegalkin1927}. The supports of the monomials in the ANF form a finite subset of \(2^{[n]}\). Its inclusion-maximal elements form an antichain and encode the top-level algebra of \(f\).

\emph{Boolean derivatives}, systematized by Thayse~\cite{thayse1981}, offer another perspective. The first-order derivative $D_i f$ measures sensitivity of $f$ to flipping $x_i$, while $D_S f$ tests simultaneous sensitivity to all of $S$. Applications include test generation~\cite{sellers1968}, formal verification~\cite{wegener1987}, and circuit complexity~\cite{jukna2012}.

The main equivalence connecting these two objects does not appear to be
widely emphasized in the literature:
\[
  D_S f \equiv 1 \quad\iff\quad
  S \text{ is a maximal ANF monomial support of } f.
\]
The condition $D_S f \equiv 1$ is a \emph{parity certificate} for
satisfiability, it guarantees that every $S$-dimensional subcube contains an
odd number of satisfying assignments, hence at least one. We call $S$ a
\emph{constant derivative certificate}. Whenever such a nonempty certificate
exists, it gives stronger evidence for satisfiability than a single point
witness: the subcube-parity condition holds universally over all
$2^{n - \abs{S}}$ base vectors $b$, whereas a point witness is a single
evaluation. We define $\Delta_\partial(f)$ as the minimum cardinality of a
constant derivative certificate.

We study the decision problem $\BCD$: does $f$ admit a constant derivative
certificate of size $\le k$? Its natural $\exists\forall\oplus$ formulation
separates the analysis into three cases: fixed $k$, where the problem is
$\coNP$-complete; $k=n$, where it is $\NP$-complete; and variable $k$, where it
lies in $\SigTwo$.

\paragraph{Related work.}
Boolean derivatives appear in Boolean difference calculus, including
Thayse's monograph~\cite{thayse1981}, and in applications such as test
generation~\cite{sellers1968}, formal verification~\cite{wegener1987}, and
symmetric cryptography~\cite{carlet2010}. Polynomial representations of
Boolean functions over finite fields are foundational in algebraic approaches to circuit lower bounds, including the Razborov--Smolensky
method~\cite{razborov1987,smolensky1987}. Toda's theorem~\cite{toda1991}
shows the uses of counting and parity in complexity theory. The
polynomial-time hierarchy originates with Stockmeyer~\cite{stockmeyer1976};
parity-oracle variants are related to work of Papadimitriou and
Zachos~\cite{papadimitriou1983}.

\paragraph{Organization.}
Section~\ref{sec:background} recalls Boolean derivatives and ANF.
Section~\ref{sec:correspondence} proves the derivative--ANF correspondence.
Section~\ref{sec:certificates} introduces derivative certificate depth.
Sections~\ref{sec:examples}--\ref{sec:sigma2} establish the complexity results.

\section{Preliminaries}
\label{sec:background}

Throughout, $[n] \coloneqq \{1,\ldots,n\}$ and $f\colon\{0,1\}^n\to\{0,1\}$.
All arithmetic on Boolean values is over $\F_2$ unless otherwise stated.

\subsection*{Boolean derivatives}

\begin{definition}
The \emph{Boolean derivative} of $f$ with respect to $x_i$ is
\[
  D_i f \;\coloneqq\; f\big|_{x_i = 0} \oplus f\big|_{x_i = 1}.
\]
For $S \subseteq [n]$, the \emph{higher-order derivative} is
$D_S f \coloneqq \bigl(\prod_{i \in S} D_i\bigr) f$,
where the product denotes operator composition. We set $D_\emptyset f \coloneqq f$.
\end{definition}

The operators $D_i$ pairwise commute over $\F_2$: for any $i \ne j$ and
$x \in \{0,1\}^n$, writing $x^{(T)}$ for $x$ with bits in $T$ flipped,
\[
  D_i(D_j f)(x)
  = f(x) \oplus f(x^{(i)}) \oplus f(x^{(j)}) \oplus f(x^{(ij)})
  = D_j(D_i f)(x).
\]
Since the operators \(D_i\) commute, \(D_S f\) is independent of the order in
which the variables of \(S\) are differentiated. Equivalently, if the variables
outside \(S\) are fixed to \(b\in\{0,1\}^{[n]\setminus S}\), then
\begin{equation}
\label{eq:subcube-parity}
  D_S f(b)
  =
  \bigoplus_{u\in\{0,1\}^S} f(b,u).
\end{equation}
Thus \(D_S f(b)\) is the parity of \(f\) on the \(S\)-dimensional coordinate
subcube obtained by varying the variables in \(S\) and fixing all others
according to \(b\).

\subsection*{Algebraic normal form}

Every Boolean function has a unique \emph{algebraic normal form}:
\[
  f(x) \;=\; \bigoplus_{T \subseteq [n]} c_T \prod_{i \in T} x_i,
  \qquad c_T \in \{0,1\},
\]
with coefficients determined by M\"{o}bius inversion on the Boolean lattice:
$c_T = \bigoplus_{R \subseteq T} f(\mathbf{1}_R)$, where $\mathbf{1}_R$ denotes
the indicator vector of $R \subseteq [n]$~\cite{jukna2012}. We write
\[
  \operatorname{supp}(f) \;\coloneqq\; \{T \subseteq [n] : c_T = 1\}
\]
for the \emph{monomial support} of $f$.

\begin{definition}
A set $T \in \operatorname{supp}(f)$ is a \emph{maximal ANF monomial support}
of $f$ if no proper superset $T' \supsetneq T$ lies in $\operatorname{supp}(f)$.
\end{definition}

\section{The Derivative--ANF Correspondence}
\label{sec:correspondence}

\begin{proposition}[Derivative formula]
\label{prop:deriv-formula}
For every $f$ and every $S \subseteq [n]$,
\begin{equation}
\label{eq:deriv-anf}
  D_S f \;=\; \bigoplus_{\substack{T \subseteq [n] \\ T \supseteq S}} c_T
  \prod_{i \in T \setminus S} x_i.
\end{equation}
\end{proposition}

\begin{proof}
By $\F_2$-linearity of $D_S$, it suffices to verify~\eqref{eq:deriv-anf} on
each monomial $m_T \coloneqq \prod_{i \in T} x_i$.

\textit{Case $S \not\subseteq T$.}
Pick $j \in S \setminus T$. Since $m_T$ does not depend on $x_j$, we have
$D_j m_T = m_T \oplus m_T = 0$, and therefore $D_S m_T = 0$. This is consistent
with~\eqref{eq:deriv-anf}, which contributes no term from $T$ (as $T \not\supseteq S$).

\textit{Case $S \subseteq T$.}
For each $j \in S$:
\[
  D_j\!\left(\prod_{i \in T} x_i\right)
  = \prod_{i \in T} x_i\Big|_{x_j=0}
    \oplus \prod_{i \in T} x_i\Big|_{x_j=1}
  = 0 \oplus \prod_{i \in T \setminus \{j\}} x_i
  = \prod_{i \in T \setminus \{j\}} x_i.
\]
Applying each $D_j$ for $j \in S$ iteratively (in any order, by commutativity)
gives $D_S m_T = \prod_{i \in T \setminus S} x_i$, which matches the $T$-term
in~\eqref{eq:deriv-anf}. Summing over $T \supseteq S$ with $c_T = 1$ completes
the proof.
\end{proof}

\begin{theorem}[Derivative--ANF correspondence]
\label{thm:correspondence}
$D_S f \equiv 1$ if and only if $S$ is a maximal ANF monomial support of $f$.
\end{theorem}

\begin{proof}
Decompose~\eqref{eq:deriv-anf} by separating the \(T=S\) term:
\[
  D_S f
  =
  c_S
  +
  \bigoplus_{\substack{T \supsetneq S \\ T \in \operatorname{supp}(f)}}
  \prod_{i\in T\setminus S}x_i .
\]
The first term is constant, while every term with \(T\supsetneq S\) is a
nonconstant multilinear monomial in the remaining variables. By uniqueness of
the ANF on the variables \([n]\setminus S\), this derivative is identically
\(1\) if and only if
\[
  c_S=1
  \qquad\text{and}\qquad
  c_T=0\text{ for every }T\supsetneq S.
\]
The first condition says that the monomial with support \(S\) appears in the
ANF of \(f\). The second says that no monomial whose support properly contains
\(S\) appears in the ANF of \(f\). Thus \(S\) is maximal among the ANF supports
under set inclusion.
\end{proof}

\begin{remark}
Equation~\eqref{eq:deriv-anf} shows that $D_S$ acts on the ANF by
\emph{projecting onto monomials containing $S$, then cancelling the
$S$-variables}. This is the Boolean analogue of partial differentiation in
$\F_2[x_1,\ldots,x_n]$.
\end{remark}

\section{Constant Derivative Certificates}
\label{sec:certificates}

\begin{definition}
\label{def:cert}
A nonempty set $S \subseteq [n]$ is a \emph{constant derivative certificate}
for $f$ if $D_S f \equiv 1$. The \emph{derivative certificate depth} of $f$ is
\[
  \Delta_\partial(f) \;\coloneqq\;
  \min\bigl\{\abs{S} : \varnothing \ne S \subseteq [n],\; D_S f \equiv 1\bigr\},
\]
with $\Delta_\partial(f) = \infty$ if no such $S$ exists. By
Theorem~\ref{thm:correspondence}, $\Delta_\partial(f)$ equals the minimum
degree of a nonempty maximal element of $\operatorname{supp}(f)$.
\end{definition}

\begin{proposition}
\label{prop:finite}
$\Delta_\partial(f) < \infty$ if and only if $f$ is nonconstant.
\end{proposition}

\begin{proof}
If $f \equiv 0$: then $\operatorname{supp}(f) = \varnothing$, so no maximal
element exists and $\Delta_\partial(f) = \infty$.

If $f \equiv 1$: then $c_\varnothing = 1$ and $c_T = 0$ for all $T \ne \varnothing$
(since Möbius inversion gives $c_\varnothing = f(\mathbf{0}) = 1$ and
$c_{\{i\}} = f(\mathbf{0}) \oplus f(\mathbf{e}_i) = 1 \oplus 1 = 0$, etc.).
So $\operatorname{supp}(f) = \{\varnothing\}$, whose unique element is empty.
Since Definition~\ref{def:cert} requires a \emph{nonempty} certificate,
$\Delta_\partial(f) = \infty$.

If $f$ is nonconstant: $\operatorname{supp}(f)$ contains at least one nonempty
set. Since $\operatorname{supp}(f)$ is a finite nonempty poset, it has a
maximal element $S$. If $S = \varnothing$, then every proper superset satisfies
$c_{T} = 0$, and $c_\varnothing = 1$; but then $f \equiv 1$ by Möbius inversion,
contradicting nonconstancy. So $S \ne \varnothing$, giving $\Delta_\partial(f)
\le \abs{S} < \infty$.
\end{proof}

\begin{remark}
The proposition shows $\Delta_\partial(f) = \infty$ for \emph{both} constant
functions, even though $f \equiv 1$ is satisfiable. The parameter
$\Delta_\partial$ does not measure satisfiability.
\end{remark}

\section{Examples}
\label{sec:examples}

\begin{example}[Linear functions]
If $f = x_{i_1} \oplus \cdots \oplus x_{i_k}$, the ANF consists of $k$
degree-one monomials $\{i_1\},\ldots,\{i_k\}$, each of which is maximal, since no higher-degree monomial appears. Hence $\Delta_\partial(f) = 1$.
\end{example}

\begin{example}[Pure conjunction]
$f = x_1 \cdots x_k$ has a single ANF monomial $[k]$, which is trivially
maximal. Hence $\Delta_\partial(f) = k$, showing that the parameter can equal the
degree of the unique maximal ANF monomial.
\end{example}

\begin{example}[Majority on three variables]
$f = x_1 x_2 \oplus x_1 x_3 \oplus x_2 x_3$. M\"{o}bius inversion gives
$c_{\{1,2,3\}} = \bigoplus_{R \subseteq \{1,2,3\}} f(\mathbf{1}_R) = 0$, so the
degree-three monomial is absent. All three degree-two monomials are present and
maximal. Hence $\Delta_\partial(f) = 2$.
\end{example}

\begin{example}[Constant functions]
$\Delta_\partial(f) = \infty$ for $f \equiv 0$ and $f \equiv 1$ by
Proposition~\ref{prop:finite}.
\end{example}

\section{The Decision Problem}
\label{sec:problem}

\begin{problem}[\textsc{Bounded Constant Derivative} ($\BCD$)]
\label{prob:bcd}
Given a Boolean circuit $f(x_1,\ldots,x_n)$ and an integer $1 \le k \le n$,
decide whether $\Delta_\partial(f) \le k$.
\end{problem}

By Theorem~\ref{thm:correspondence}, $\BCD$ asks whether $\operatorname{supp}(f)$
has a maximal element of cardinality $\le k$. Expanding via~\eqref{eq:subcube-parity}:
\begin{equation}
\label{eq:quantifier-structure}
  \exists\, S \subseteq [n],\; 0 < \abs{S} \le k
  \quad\forall\, b \in \{0,1\}^{[n]\setminus S}
  \quad \bigoplus_{u \in \{0,1\}^S} f(b, u) = 1. \tag{$*$}
\end{equation}
The $\exists\forall\oplus$ quantifier structure of $(*)$ controls the complexity of $\BCD$. The three layers work differently with $k$: at fixed $k$ the parity layer ($\oplus$) collapses to polynomial time; at $k = n$ the $\forall$-layer ranges over the single empty tuple; for general $k$ all three are utilized.

\section{Fixed-Depth Complexity}
\label{sec:fixedk}

\begin{theorem}
\label{thm:conp}
For every fixed constant $K \ge 1$, the problem $\BCDK$ is $\coNP$-complete.
\end{theorem}

\begin{proof}
\textbf{\(\coNP\)-membership.}
We show \(\overline{\BCDK}\in\NP\). An instance \((f,K)\) is a no-instance
iff every nonempty \(S\subseteq[n]\) with \(\abs{S}\le K\) satisfies
\(D_S f\not\equiv 1\), i.e., there exists
\(b_S\in\{0,1\}^{[n]\setminus S}\) with \(D_S f(b_S)=0\).
A nondeterministic machine guesses the family \(\{b_S\}_S\) and verifies each
condition
\[
  \bigoplus_{u\in\{0,1\}^S} f(b_S,u)=0
\]
by evaluating \(f\) at at most \(2^K\) points. Since \(K\) is fixed and there
are \(\sum_{j=1}^K {n\choose j}=O(n^K)\) choices of \(S\), verification is
polynomial in \(n\). Hence \(\overline{\BCDK}\in\NP\), so
\(\BCDK\in\coNP\).

\textbf{$\coNP$-hardness.}
We give a polynomial-time many-one reduction from $\Taut$ to $\BCDK$. Given a
formula $\psi(y_1,\ldots,y_m)$, introduce variables $x_1,\ldots,x_K$
(disjoint from $\mathbf{y}$) and define
\[
  f(x_1,\ldots,x_K,\,y_1,\ldots,y_m)
  \;\coloneqq\; x_1 \cdots x_K \cdot \psi(y_1,\ldots,y_m).
\]
Let $X \coloneqq \{1, \ldots, K\}$ denote the indices of the $x$-variables.

\medskip
\textit{Step 1: ANF of $f$.}
Since the $x$-variables and $y$-variables are disjoint, the product distributes as
\[
  f
  = \left(\prod_{i \in X} x_i\right)
    \cdot \left(\bigoplus_{T \subseteq [m]} c_T^{\psi}\prod_{j \in T} y_j\right)
  = \bigoplus_{T \subseteq [m]} c_T^{\psi}
    \left(\prod_{i \in X} x_i\right) \left(\prod_{j \in T} y_j\right),
\]
where $c_T^\psi$ denotes the ANF coefficient of $\psi$ for the monomial $\prod_{j \in T} y_j$. For any $T \subseteq [m]$, let $T' \coloneqq \{j + K : j \in T\}$ be the corresponding indices in $f$. Hence
\[
  \operatorname{supp}(f)
  = \bigl\{ X \cup T' : T \subseteq [m],\; c_T^\psi = 1 \bigr\}.
\]
Thus every ANF monomial of $f$ contains $X$ as a subset and has degree $\ge K$.

\medskip
\textit{Step 2: Any certificate of size $\le K$ must equal $X$.}
Suppose $S \subseteq [n]$ with $\abs{S} \le K$ is a constant derivative
certificate, so $D_S f \equiv 1$. We first show $X \subseteq S$. Suppose for
contradiction that $X \not\subseteq S$, and pick any $j \in X \setminus S$.
By Proposition~\ref{prop:deriv-formula},
\[
  D_S f
  = \bigoplus_{\substack{M \in \operatorname{supp}(f) \\ M \supseteq S}}
    \prod_{i \in M \setminus S} x_i.
\]
Since every \(M \in \operatorname{supp}(f)\) contains all of \(X\), we have
\(j \in M \setminus S\) for every contributing term. Thus the variable \(x_j\)
divides every summand, giving $D_S f = x_j \cdot g$ for some Boolean polynomial $g$. Setting
$x_j = 0$ then yields $D_S f\big|_{x_j=0} \equiv 0 \ne 1$, contradicting
$D_S f \equiv 1$. Hence $X \subseteq S$. Combined with $\abs{S} \le K = \abs{X}$,
we get $S = X$.

\medskip
\textit{Step 3: $\Delta_\partial(f) \le K$ iff $\psi \equiv 1$.}
By Step~2, $\Delta_\partial(f) \le K$ iff $D_X f \equiv 1$. Applying
Proposition~\ref{prop:deriv-formula} with $S = X$:
\[
  D_X f
  = \bigoplus_{T \subseteq [m]} c_T^\psi \prod_{j \in T} y_j
  = \psi.
\]
Hence $D_X f \equiv 1$ iff $\psi \equiv 1$. The reduction is polynomial-time
and $\Taut$ is $\coNP$-complete~\cite{cook1971}.
\end{proof}

\section{The Unrestricted Case: \texorpdfstring{$k = n$}{k = n}}
\label{sec:unrestricted}

\begin{theorem}
\label{thm:np}
$\BCD$ with $k = n$ is $\NP$-complete.
\end{theorem}

\begin{proof}
When $k = n$, the condition $\Delta_\partial(f) \le n$ is equivalent, by
Proposition~\ref{prop:finite}, to $f$ being nonconstant (since every nonempty
monomial has degree $\le n$).

\textbf{$\NP$-membership.}
A pair of assignments $a, b \in \{0,1\}^n$ with $f(a) \ne f(b)$ is a
polynomial-time verifiable witness for nonconstancy.

\textbf{$\NP$-hardness.}
We reduce $\textsc{Sat}$ to nonconstancy. Given $\varphi(y_1,\ldots,y_m)$,
introduce a variable $z$ and define
$f(y_1,\ldots,y_m,z) \coloneqq z \cdot \varphi(y_1,\ldots,y_m)$.

If $\varphi$ is unsatisfiable, then $\varphi \equiv 0$ as a Boolean function,
so $f \equiv 0$ is constant.
If $\varphi$ is satisfiable with witness $a \in \{0,1\}^m$, then
$f(a, 0) = 0 \ne 1 = f(a, 1)$, so $f$ is nonconstant.
The reduction runs in polynomial time.
\end{proof}

\begin{remark}
The transition from $\coNP$ to $\NP$ as $k$ grows is quite interesting. In $(*)$, the $\forall$-layer ranges over $2^{n-k}$ base vectors: at fixed $k$ this is exponential, making the $\forall$-layer the bottleneck; at $k = n$ it ranges over $\{0,1\}^0 = \{()\}$, a single empty tuple, making it vacuous. As $k$ increases, the $\forall$-layer collapses and the complexity seems to go from $\coNP$ toward $\NP$.
\end{remark}

\section{The Variable-\texorpdfstring{$k$}{k} Case and
         \texorpdfstring{$\SigTwo$}{Sigma2oP}}
\label{sec:sigma2}

We define $\SigTwo$ as the class of decision problems of the form
\[
  x \in L \;\iff\;
  \exists\, S \in \{0,1\}^{p(\abs{x})}
  \;\forall\, b \in \{0,1\}^{q(\abs{x})}
  \;\bigoplus_{u \in \{0,1\}^{r(\abs{x})}} C(x, S, b, u) = 1
\]
for some polynomial-time computable Boolean circuit $C$ and polynomials
$p, q, r$. This is the parity-oracle analogue of the second existential--universal level;
equivalently, one may view it as \(\NP^{\PiOne}\), where
\(\PiOne = \coNP^{\oP}\), as considered by Papadimitriou and
Zachos~\cite{papadimitriou1983}.

\begin{proposition}
\label{prop:sigma2}
The general problem $\BCD$ (variable $k$) lies in $\SigTwo$ and is both
$\NP$-hard and $\coNP$-hard under polynomial-time many-one reductions.
\end{proposition}

\begin{proof}
\textbf{Hardness.}
Immediate from Theorems~\ref{thm:conp} and~\ref{thm:np}: the fixed-$K$
and $k = n$ specializations of $\BCD$ are $\coNP$-hard and $\NP$-hard,
respectively.

\textbf{\(\SigTwo\)-membership.}
The quantifier structure~\eqref{eq:quantifier-structure} directly witnesses
membership. Equivalently, a \(\NP^{\PiOne}\) algorithm proceeds as follows.
First, nondeterministically guess \(S\subseteq[n]\) and check in polynomial time
that \(0<\abs{S}\le k\). For this fixed \(S\), the required verification
condition is
\[
  \forall b\in\{0,1\}^{[n]\setminus S}
  \quad
  \bigoplus_{u\in\{0,1\}^{S}} f(b,u)=1.
\]
For fixed \(S\) and \(b\), the inner parity predicate is decidable by an
\(\oP\) oracle: it asks whether the number of \(u\in\{0,1\}^S\) satisfying
\(f(b,u)=1\) is odd. Hence the universal verification condition lies in
\(\coNP^{\oP}=\PiOne\). The outer nondeterministic machine accepts exactly when
this \(\PiOne\)-oracle verifies the condition.

Formally, encode \(S\) by its length-\(n\) characteristic vector. Let \(b\) and
\(u\) range over \(\{0,1\}^n\), with the coordinates of \(b\) inside \(S\)
ignored. Define
\[
  x_i(S,b,u)=
  \begin{cases}
    u_i, & i\in S,\\
    b_i, & i\notin S.
  \end{cases}
\]
Now define the polynomial-time predicate
\[
  C(f,S,b,u)=
  \begin{cases}
    f(x(S,b,u)), & \text{if } u_i=0 \text{ for every } i\notin S,\\
    0, & \text{otherwise.}
  \end{cases}
\]
Then
\[
  \bigoplus_{u\in\{0,1\}^n} C(f,S,b,u)
  =
  \bigoplus_{u\in\{0,1\}^S} f(b,u).
\]
Thus the variable-length parity over \(\{0,1\}^S\) can be represented as a
fixed \(n\)-bit parity predicate, placing the verification condition in
\(\PiOne=\coNP^{\oP}\). Since the outer machine nondeterministically guesses
\(S\) and then invokes this \(\PiOne\) verification condition, we obtain $\BCD\in\NP^{\PiOne}=\SigTwo.$
\end{proof}

At fixed $k = K$, the inner parity $\bigoplus_{u \in \{0,1\}^K} f(b,u)$ is computable in polynomial time because $K$ is constant. Thus the parity-oracle layer disappears, consistent with the $\coNP$-completeness result above.
At $k = n$, the $\forall$-layer is vacuous and the problem reduces to pure $\NP$, consistent with Theorem~\ref{thm:np}.

The class \(\SigTwo\) contains the ordinary \(\Sigma_2^P\) as a special case:
take the inner parity range to have size one and let the circuit compute an
ordinary polynomial-time predicate. It is also contained in \(\PSPACE\), since
a polynomial-space machine can enumerate the existential witness, universally
check all \(b\), and compute each inner parity by enumeration. Thus we use only
the conservative containments
\[
  \NP,\;\coNP
  \;\subseteq\; \Sigma_2^P
  \;\subseteq\; \SigTwo
  \;\subseteq\; \PSPACE.
\]
We do not rely on any strict separations among these classes.

We conjecture that $\BCD$ is $\SigTwo$-complete. A proof would
reduce a $\exists S\,\forall b\,[\bigoplus_u C(S,b,u)=1]$-complete
problem to $\BCD$ by constructing a circuit $f$ for which $\bigoplus_u f(b,u)$
simulates $\bigoplus_u C(S,b,u)$ under the identification $S \leftrightarrow
\text{existential witness}$, $b \leftrightarrow \text{universal input}$; we
leave this open.

The complexity of $\BCD$ as $k$ varies is summarized in
Table~\ref{tab:complexity}.

\begin{table}[h]
\centering
\renewcommand{\arraystretch}{1.45}
\begin{tabular}{
  >{\centering\arraybackslash}m{3.5cm}
  >{\centering\arraybackslash}m{4.3cm}
  m{5.0cm}}
\toprule
$k$ & Established complexity & Active quantifier layers \\
\midrule
Fixed constant $K$ & $\coNP$-complete
  & $\forall$-layer dominates; the $\oplus$-layer is polynomial-time \\[3pt]
Variable $k$ & In $\SigTwo$; $\NP$-hard and $\coNP$-hard
  & $\exists$, $\forall$, and $\oplus$ layers can all be active; completeness is left open \\[3pt]
$k = n$ & $\NP$-complete
  & $\forall$-layer is vacuous; the problem reduces to nonconstancy \\
\bottomrule
\end{tabular}
\caption{Established complexity of $\BCD$ as $k$ varies.}
\label{tab:complexity}
\end{table}

\subsection*{The Derivative Lattice}

The family of all higher-order derivatives of $f$ can be organized as the
\emph{derivative lattice}: nodes are the derivatives $D_S f$ for $S \subseteq
[n]$, ordered by inclusion of $S$, with an edge from $D_S f$ to $D_{S \cup
\{i\}} f$ whenever $i \notin S$. Level $k$ contains exactly $\binom{n}{k}$
nodes, the derivatives of order $k$. This is actually the Hasse diagram of
the Boolean lattice $2^{[n]}$, shown by the fact that the operators $D_i$
commute: the node reached by applying $D_i$ then $D_j$ is the same as the node
reached by applying $D_j$ then $D_i$.

\begin{figure}[h]
\centering
\begin{tikzpicture}[
  every node/.style={
    draw, rounded corners=3pt,
    minimum width=2.0cm, minimum height=0.55cm,
    font=\small, align=center
  },
  accept/.style={fill=black!12, draw=black, very thick},
  edge/.style={-{Stealth[length=5pt]}, gray!70, line width=0.7pt},
  level/.style={draw=none, fill=none, font=\footnotesize\itshape,
                minimum width=0pt, minimum height=0pt}
]

\node (f)  at (0, 0)    {$D_{\varnothing}f = f$};

\node (D1) at (-3.5,-1.7) {$D_{\{1\}}f$};
\node (D2) at ( 0,  -1.7) {$D_{\{2\}}f$};
\node (D3) at ( 3.5,-1.7) {$D_{\{3\}}f$};

\node [accept] (D12) at (-3.5,-3.4) {$D_{\{1,2\}}f \equiv 1$};
\node          (D13) at ( 0,  -3.4) {$D_{\{1,3\}}f$};
\node          (D23) at ( 3.5,-3.4) {$D_{\{2,3\}}f$};

\node (D123) at (0,-5.1) {$D_{\{1,2,3\}}f$};

\draw[edge] (f) -- (D1);
\draw[edge] (f) -- (D2);
\draw[edge] (f) -- (D3);

\draw[edge] (D1) -- (D12);
\draw[edge] (D1) -- (D13);
\draw[edge] (D2) -- (D12);
\draw[edge] (D2) -- (D23);
\draw[edge] (D3) -- (D13);
\draw[edge] (D3) -- (D23);

\draw[edge] (D12) -- (D123);
\draw[edge] (D13) -- (D123);
\draw[edge] (D23) -- (D123);

\node[level] at ( 6.0,  0.0) {level $0$};
\node[level] at ( 6.0, -1.7) {level $1$};
\node[level] at ( 6.0, -3.4) {level $2$};
\node[level] at ( 6.0, -5.1) {level $3$};

\end{tikzpicture}
\caption{The derivative lattice for $n = 3$. Each node $D_S f$ sits at
level $\abs{S}$; edges go from $D_S f$ to $D_{S \cup \{i\}} f$ for $i \notin
S$. A node is \emph{accepting} (shaded, bold border) when $D_S f \equiv 1$,
equivalently when $S$ is a maximal ANF monomial support of $f$. In the example
shown, an accepting node first appears at level $2$, so
$\Delta_\partial(f)=2$.}
\label{fig:lattice}
\end{figure}
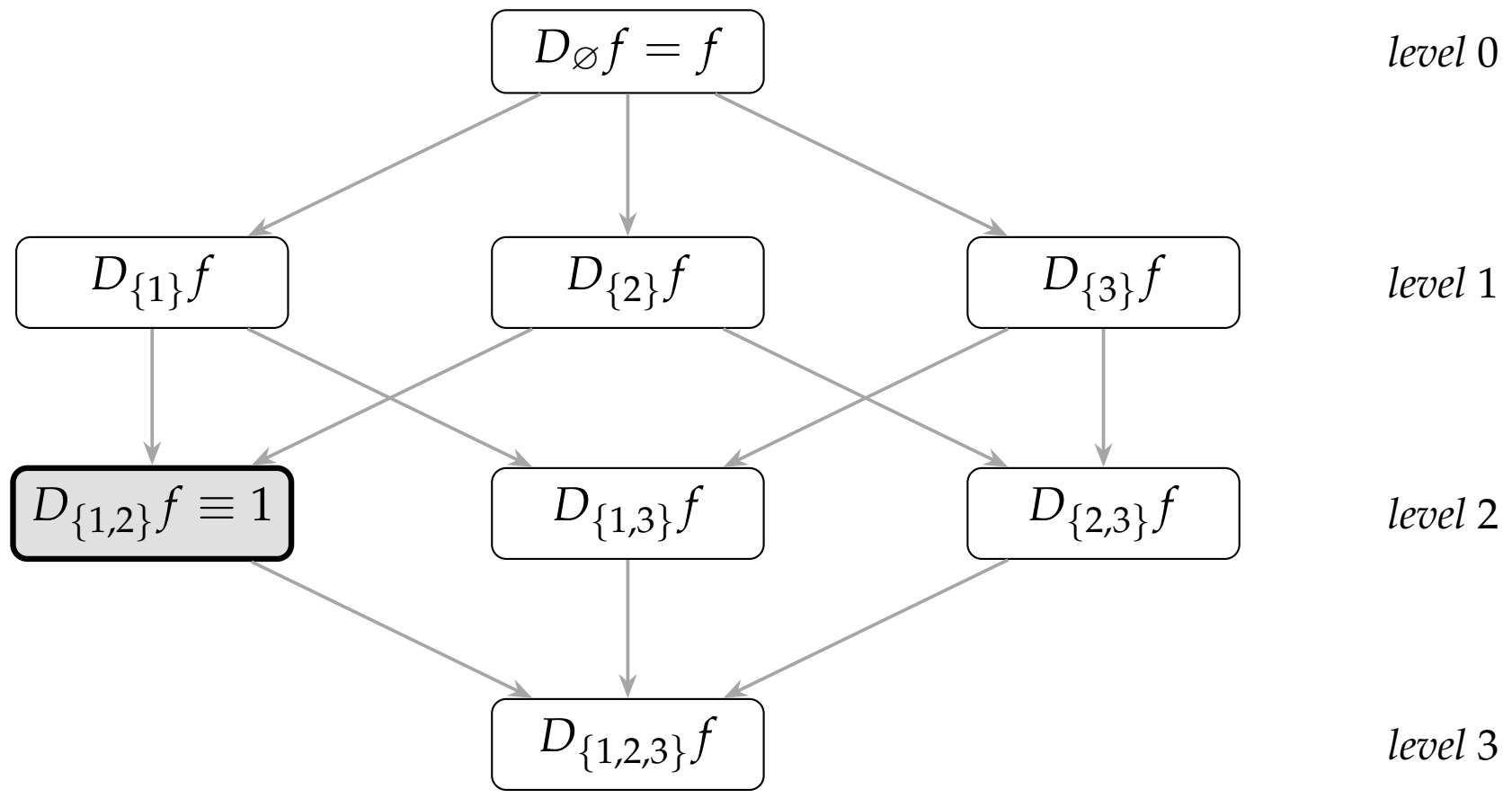

In this picture, Theorem~\ref{thm:correspondence} says that a node $D_S f$ is
\emph{accepting} (identically $1$) if and only if $S$ is a maximal ANF monomial
support of $f$. The derivative certificate depth $\Delta_\partial(f)$ is then
the \emph{minimum level} containing an accepting node. Figure~\ref{fig:lattice}
illustrates this for $n = 3$.

\bibliographystyle{plain}

\end{document}